\documentclass[english]{article}
\usepackage[T1]{fontenc}
\usepackage[latin9]{inputenc}
\usepackage[dvipsnames]{xcolor}
\usepackage{array}
\usepackage{multirow}
\usepackage{amstext}
\usepackage{amssymb}
\usepackage{graphicx}
\usepackage{setspace}
\usepackage{esint}

\makeatletter

\providecommand{\tabularnewline}{\\}

\usepackage{amsfonts,amssymb}
\usepackage{latexsym}
\usepackage{epsfig}
\usepackage{cite}
\usepackage{graphicx}
\usepackage[colorlinks,linkcolor=blue]{hyperref}
\newcommand{\be}{\begin{equation}}
\newcommand{\ee}{\end{equation}}

\makeatother

\usepackage{babel}
\begin{document}
{}~ \hfill\vbox{\hbox{CTP-SCU/2023006}}\break \vskip 3.0cm 
\centerline{\Large \bf Timelike  entanglement entropy in $\text{dS}_3/\text{CFT}_2$}
\vspace*{10.0ex} \centerline{\large Xin Jiang$^a$, Peng Wang$^a$, Houwen Wu$^{a,b}$ and Haitang Yang$^a$} \vspace*{7.0ex} \vspace*{4.0ex} \centerline{\large \it $^a$College of Physics} \centerline{\large \it Sichuan University} \centerline{\large \it Chengdu, 610065, China} \vspace*{1.0ex} \vspace*{4.0ex} \centerline{\large \it $^b$DAMTP, Centre for Mathematical Sciences} \centerline{\large \it University of Cambridge} \centerline{\large \it Cambridge, CB3 0WA, UK} \vspace*{1.0ex} \vspace*{4.0ex}
\centerline{xjang@stu.scu.edu.cn, pengw@scu.edu.cn, hw598@damtp.cam.ac.uk, hyanga@scu.edu.cn} \vspace*{10.0ex} \centerline{\bf Abstract} \bigskip \smallskip 

In the context of  dS$_3$/CFT$_2$, we propose a timelike entanglement entropy defined by the renormalization group flow. 
This  timelike entanglement entropy is calculated in CFT by using the  Callan-Symanzik equation. 
We find an exact match between this entanglement entropy 
and the length of a timelike geodesic connecting two different spacelike surfaces in dS$_3$.
The counterpart of this entanglement entropy in AdS$_3$ is a spacelike one, also induced by RG flow and extends all the 
way into the bulk of AdS$_3$. 
As a result, in both AdS$_3$/CFT$_2$ and  dS$_3$/CFT$_2$, there exist exactly three 
entanglement entropies, providing precisely sufficient information to reconstruct the three-dimensional bulk geometry.

\vfill 
\eject
\baselineskip=16pt
\vspace*{10.0ex}


{}

{}

\section{Introduction}

The gauge/gravity correspondence \cite{Maldacena:1997re,Witten:1998qj,Aharony:1999ti,Strominger:2001pn,Bousso:2002ju}
states that a $(d+1)$-dimensional gravitational theory is equivalent
to a $d$-dimensional conformal field theory (CFT), which provides
a computational holographic realization. 
One of the central concerns in the duality is to  extract the bulk geometry, kinematics and even dynamics, 
from the CFT quantities. Since Ryu and Takayanagi identified the holographic entanglement entropy (EE) of CFTs 
with the lengths of the geodesics anchored on the   conformal boundary of the bulk geometry in \cite{Ryu:2006bv,Ryu:2006ef}, it is believed that EE  should play a major role in constructing the bulk geometry.

However, recent work has suggested that the traditional {\it spacelike}
EE  does not fully capture the  entangling properties of CFTs. 
In Ref. \cite{Wang:2018jva}, we observed that there exists a severe inconsistency in the corrected EE of 
the $T\bar{T}$ deformed version of the $\text{AdS}_{3}$/$\text{CFT}_{2}$
correspondence.  To resolve the inconsistency, we proposed that, 
in addition to the traditional spacelike EE,  
a {\it timelike} EE must be introduced. Remarkably,
such a timelike EE has been explicitly addressed
in the $\text{AdS}$/$\text{CFT}$ context 
\cite{Doi:2022iyj,Narayan:2022afv,Li:2022tsv,Doi:2023zaf,Jiang:2023ffu,Narayan:2023ebn}
recently. Unlike the traditional spacelike EE, which measures
entangling between spacelike subsystems, the timelike
EE reflects the entangling between timelike intervals. 
Since the $2d$ conformal boundary of $\text{AdS}_{3}$ is  Minkovski, 
it looks natural that both spacelike and
timelike EEs exist in the $\text{AdS}_{3}$/$\text{CFT}_{2}$.

%
%


The timelike EE in the context of the $\text{dS}_{3}$/$\text{CFT}_{2}$
correspondence \cite{Strominger:2001pn,Maldacena:2002vr,Hikida:2021ese,Hikida:2022ltr}
is perhaps more intriguing. Naively, it appears that    $\text{dS}_{3}$/$\text{CFT}_{2}$
leaves no room for  a timelike EE.
To see this clearly, let us consider the  $\text{dS}_{3}$ in the planar coordinate,
\begin{equation}
ds_{\rm dS}^{2}=-dt^{2}+e^{2t/\ell_{\text{dS}}}\,\left(dx^{2}+dy^{2}\right),\label{eq:flat_metric}
\end{equation}
where $t\in(-\infty,\infty)$ and  $\ell_{\text{dS}}$ is the dS radius.
The future (or equivalently the past) boundary $\mathcal{I}^{+}$ as $t\rightarrow+\infty$,
where the dual $\text{CFT}_{2}$ lives,
is obviously a Euclidean plane that  has two spatial directions. 
On the other hand, the  $\text{AdS}_{3}$ in the planar coordinate reads,
\begin{equation}
ds_{\rm AdS}^{2}=d\xi^{2}+e^{2\xi/\ell_{\rm AdS}}\,\left(-dt^{2}+dx^{2}\right),\label{eq: AdS_Planar}
\end{equation}
which is related to the familiar Poincare coordinate with $\xi/\ell_{\rm AdS} \to - \log [z/\ell_{\rm AdS}]$.
It is then easy to see that, the double Wick rotation, 
\begin{equation}
\xi \to it,\quad t\to iy\quad  \ell_{\rm AdS}\to i\ell_{\rm dS}, \quad x\to x,\label{eq: Wick rotation}
\end{equation}
a usual operation from $\text{AdS}_{3}$ to $\text{dS}_{3}$,
transforms both the  spacelike and timelike EEs in $\text{AdS}_{3}$/$\text{CFT}_{2}$
to spacelike EEs along the two directions in
$\text{dS}_{3}$/$\text{CFT}_{2}$.

\begin{figure}
\begin{centering}
\includegraphics[width=0.5\textwidth]{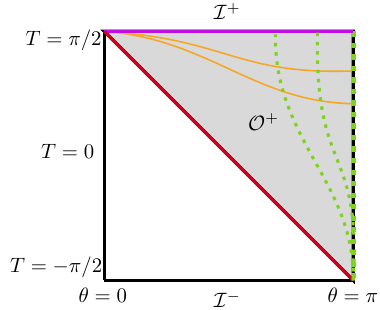}\label{fig:Penrose}
\par\end{centering}
\caption{Penrose diagram of de Sitter spacetime.  $\mathcal{I}^{\pm}$ is the {\it global} past and future spheres. The two vertical 
boundaries $\theta=0,\pi$ are the north pole and south pole respectively. Each point in the interior
represents an $\mathbb{S}^{1}$. A horizontal slice is an $\mathbb{S}^{2}$. 
The planar coordinate (\ref{eq:flat_metric}) covers the shadow
region  $\mathcal{O}^{+}$, comprising the causal future of the south pole. The green dashed lines are constant $r=\sqrt{x^{2}+y^{2}}$. Orange lines
of constant $t$ are shown. 
The violet line $\mathcal{I}^{+}$  denotes the future boundary $t\to \infty$. 
The red line indicates the past horizon  $t\to -\infty$.}
\end{figure}

Remarkably,  the above double Wick rotation  already gives some hints 
about the existence of a  timelike EE in $\text{dS}_{3}$/$\text{CFT}_{2}$.
The purpose of this paper is to introduce a nontrivial
timelike EE in the $\text{dS}_{3}$/$\text{CFT}_{2}$
context, which might be interpreted as the entanglement between the
future and the past. 
To this end, it is of help to  draw the Penrose diagram of  $\text{dS}_{3}$ in Figure (\ref{fig:Penrose}) by using 
the conformal coordinate,
\begin{equation}
ds_{\rm dS}^{2}= \frac{1}{\cos^2 T} (-dT^2+ d\theta^2 + \sin^2\theta d\phi^2),\label{eq: dS_Conformal}
\end{equation}
with  $-\pi/2 < T < \pi/2$. In the figure, $\mathcal{I}^{\pm}$ are the {\it global} past and future spheres. The two vertical 
boundaries $\theta=0,\pi$ are the north pole and south pole respectively. Each point in the interior
represents an $\mathbb{S}^{1}$. A horizontal slice is an $\mathbb{S}^{2}$. 
The planar coordinate (\ref{eq:flat_metric}) covers the shadow
region  $\mathcal{O}^{+}$, comprising the causal future of the south pole. The green dashed lines are constant $r=\sqrt{x^{2}+y^{2}}$. Orange lines
of constant $t$ are shown. 
The violet line $\mathcal{I}^{+}$  denotes the future boundary $t\to \infty$. 
The red line indicates the past horizon  $t\to -\infty$.

In $\text{dS}_{3}$/$\text{CFT}_{2}$,
it is sufficient to consider  the inflation patch $\mathcal{O}^{+}$,
which consists of a collection of flat spacelike slices. Ref.\cite{Strominger:2001pn}
suggested that time evolution along these slices in dS$_3$ is equivalent to
scale transformations of the dual $\text{CFT}_2$. In other words, time
evolution is the inverse renormalization group (RG) flow \cite{Strominger:2001gp},
and the temporal dimension in the bulk can be interpreted as the renormalization
scale $\mu$ of the CFT, where  the future boundary $\mathcal{I}^{+}$
corresponds to the ultraviolet (UV) region of the CFT and the past
horizon (red line with $t\to -\infty$) corresponds to the infrared (IR) region of the CFT. 
Guided by these facts, we will show that in $\text{dS}_{3}$/$\text{CFT}_{2}$,
a timelike EE can be defined through the RG flow equation, which turns out to be
$S_{A}=-\frac{\text{i}\,c_{\text{dS}}}{6}\log\frac{\xi}{\epsilon}$, where $c_{\text{dS}}$  is 
the central charge, $\epsilon$ is the UV cutoff and $\xi$ is an IR-like cutoff (correlation length). 
Since this RG flow induced EE  involves both IR and UV cutoffs, 
it is convenient to refer to it as IR  EE. 
The usual EE involving only UV cutoff is 
then called UV EE to distinguish between the two.
Based on the hypothesis that the RG flow of the non-unitary  CFT  is dual to cosmological time
evolution in dS, we will demonstrate that the IR EE is dual to a timelike geodesic connecting 
spacelike surfaces at different times in the bulk of dS$_3$.
Our derivations do not require unitarity of the CFT,
which is consistent with non-unitary CFT duals of de Sitter space
\cite{Strominger:2001pn,Maldacena:2002vr,Hikida:2021ese,Hikida:2022ltr}. 

In the dS/CFT correspondence, we further clarify that timelike and spacelike
EEs are intrinsically different.  It is not surprising to find that,
via an analytical continuation, the timelike IR EE in dS/CFT 
rotates to a spacelike IR EE in AdS/CFT, 
which obviously  
is associated with the emergent radial direction of AdS! 

It is then illuminating to note that, in both $\text{dS}_{3}$/$\text{CFT}_{2}$ and 
$\text{AdS}_{3}$/$\text{CFT}_{2}$,  there are exactly three EEs,
which fits precisely to reconstruct the three-dimensional bulk geometry.
This is just 
what has been studied in  Refs. \cite{Wang:2018jva,Wang:2017bym,Wang:2018vbw}.

The remainder of this paper is outlined as follows. In section 2,
we derive the IR EE by using the Callan-Symanzik equation in CFT.
In section 3, we calculate the length of a timelike geodesic which
connects two distinct spacelike boundaries and find it matches the 
IR EE perfectly. Section \ref{Sec:Conc} is for conclusion and discussions.

\section{RG flow induced entanglement entropy in QFT}

In this section,  using the Callan-Symanzik equation, we present a universal derivation of the RG flow induced EE, 
which we refer to as the IR EE, for a generic CFT. 
When applied to the context of dS/CFT, the IR EE is timelike.

The dS/CFT correspondence   is not as well understood
as the AdS/CFT correspondence. There are only limited 
examples of CFTs that are dual to dS. In addition to a four dimensional 
higher spin gravity example \cite{Anninos:2011ui},
a recent remarkable  construction has been given for the $\text{dS}_{3}/\text{CFT}_{2}$
correspondence \cite{Hikida:2021ese,Hikida:2022ltr}. 

In the canonical formalism for  gravity,
the quantum state residing on a static compact slice $\Sigma_{t}$ can be described by the
Hartle-Hawking wavefunction $\Psi_{\text{dS}}\left[\gamma\right]$, where $\gamma$ is the metric 
on $\Sigma_{t}$. The dS/CFT could be defined through the dictionary \cite{Maldacena:2002vr},
\begin{equation}
\Psi_{\text{dS}}\left[\gamma\right]=Z_{\text{CFT}}\left[\gamma\right],\quad t\rightarrow\infty
\end{equation}
where $Z_{\text{CFT}}$ is the partition function of the  $\text{CFT}_{2}$
living on $\Sigma_{\infty}$.  Since the CFTs dual to dS 
are non-unitary \cite{Strominger:2001pn,Maldacena:2002vr,Hikida:2021ese,Hikida:2022ltr},
another universal  quantity is needed  to measure the entanglement. To this end, parallel
to the EE in unitary CFT,  a complex-valued
quantity known as the pseudoentropy   is introduced in Refs. \cite{Narayan:2015vda,Sato2015,Narayan:2015oka,Narayan:2016xwq,
Nakata:2020luh,Mollabashi:2020yie,Nishioka:2021cxe,Mollabashi:2021xsd,Doi:2022iyj}. 
Dividing
the total system into two subsystems $A$ and $B$, the pseudoentropy
is defined by the von Neumann entropy, 
\begin{equation}
S_{A}=-\mathrm{Tr}\left[\tau_{A}\log\tau_{A}\right],
\label{eq: pseudoentropy}
\end{equation}
of the reduced transition matrix 
\begin{equation}
\tau_{A}=\mathrm{Tr}_{B}\left[\frac{\left|\psi\right\rangle \left\langle  
\varphi\right|}{\left\langle \varphi\mid\psi\right\rangle }\right].
\end{equation}
Here, $\left|\psi\right\rangle $ and $\left|\varphi\right\rangle $
are two different quantum states in the total Hilbert space that is
factorized as $\mathcal{H}=\mathcal{H}_{A}\otimes\mathcal{H}_{B}$.
It should be emphasized that the pseudoentropy defined by eqn. (\ref{eq: pseudoentropy}) is generic
and does not depend on the details of a particular non-unitary CFT.
As the usual EE, the pseudoentropy could also be captured by the replica method
\cite{Calabrese:2004eu,Calabrese:2009qy} in path integral formalism.
Denoting the manifold corresponding to $\left\langle \varphi\mid\psi\right\rangle $
as $\mathcal{M}_{1}$ and the manifold corresponding to $\mathrm{Tr}_{A}\left(\tau_{A}\right)^{n}$
as $\mathcal{M}_{n}$, the $n$-th pseudo R\'enyi entropy reads
\begin{equation}
S_{A}^{\left(n\right)}=\frac{1}{1-n}\log\left[\frac{Z_{\mathcal{M}_{n}}}{\left(Z_{\mathcal{M}_{1}}\right)^{n}}\right],
\label{eq:Renyi entropy}
\end{equation}
where $Z_{\mathcal{M}}$ is the partition function over the manifold $\mathcal{M}$.
The  $n$-sheeted Riemann surfaces $\mathcal{M}_{n}$ in dS is constructed in the same way as in AdS.
Taking the limit  $n\rightarrow1$ yields the pseudoentropy
\begin{equation}
S_{A}=\lim_{n\rightarrow1}\frac{1}{1-n}\log\left[\frac{Z_{\mathcal{M}_{n}}}{\left(Z_{\mathcal{M}_{1}}\right)^{n}}\right],\label{eq:entropy}
\end{equation}
which can be regarded as a well-defined EE in the
dS$_{3}$/CFT$_{2}$ context. 




Since the definitions (\ref{eq:Renyi entropy}) and  (\ref{eq:entropy}) are identical
for both usual EE and pseudoentropy, the following derivations of IR EE are 
applicable to both unitary and non-unitary CFTs. 
Consider a $2d$ generic CFT$_2$ living on a curved surface $\mathcal{M}$
with the metric $ds^{2}=\gamma_{ab}dx^{a}dx^{b}$. It is known that
\cite{Osborn:1991gm}, for a classically scale-invariant theory where
only dimensionless couplings are present, the Callan-Symanzik equation
is
\begin{equation}
\left[\mu\frac{\partial}{\partial\mu}+2\int d^{2}x\,\gamma^{ab}\frac{\delta}{\delta\gamma^{ab}}\right]\log Z_{\text{CFT}}=0,
\end{equation}
with the renormalization scale $\mu$. The $n$-th R\'enyi entropy (\ref{eq:Renyi entropy})
of the subsystem $A$ thus satisfies 
\begin{equation}
\left[\mu\frac{\partial}{\partial\mu}+2\int d^{2}x\,\gamma^{ab}\frac{\delta}{\delta\gamma^{ab}}\right]S_{A}^{(n)}=0.\label{eq:renyi}
\end{equation}
Note that the expectation value of the stress tensor is given by
\begin{equation}
\left\langle T_{a}^{a}\right\rangle =-2\frac{\gamma^{ab}}{\sqrt{\gamma}}\frac{\delta}{\delta\gamma^{ab}}\log Z_{\text{CFT}}.\label{eq:stress_tensor}
\end{equation}
So the Callan-Symanzik equation of the $n$-th R\'enyi entropy
can be rewritten as
\begin{equation}
\mu\frac{\partial}{\partial\mu}S_{A}^{(n)}=-\frac{\int_{\mathcal{M}_{n}}\left\langle T_{a}^{a}\right\rangle _{\mathcal{M}_{n}}-n\int_{\mathcal{M}_{1}}\left\langle T_{a}^{a}\right\rangle _{\mathcal{M}_{1}}}{1-n}.
\end{equation}
The central charge $c$ of the $2d$ CFT has a clear definition due
to the presence of the Weyl anomaly
\begin{equation}
\left\langle T_{a}^{a}\right\rangle =+\frac{1}{2\pi}\frac{c}{12}\mathcal{R},\label{eq:anomaly}
\end{equation}
with $\mathcal{R}$ the scalar curvature, and one has
\begin{equation}
\mu\frac{\partial}{\partial\mu}S_{A}^{(n)}=-\frac{c\,\left(\int_{\mathcal{M}_{n}}\mathcal{R}^{(n)}-n\int_{\mathcal{M}_{1}}\mathcal{R}\right)}{24\pi\left(1-n\right)}.
\end{equation}
For the $n$-sheeted Riemannian surface in the presence of conical
singularities, Ref.\cite{Fursaev:1995ef} has shown that
\begin{equation}
\int_{\mathcal{M}_{n}}\mathcal{R}^{(n)}=n\int_{\mathcal{M}_{1}}\mathcal{R}+4\pi\left(1-n\right)\int_{\Sigma}1,
\end{equation}
where $\Sigma$ is the entangling surface. In our case here, $\left(\int_{\Sigma=\partial A}1\right)=\mathcal{A}$,
which is the number of the boundary points of the subsystem $A$.
Therefore, the Callan-Symanzik equation of the $n$-th R\'enyi entropy
could be simplified as
\begin{equation}
\mu\frac{\partial}{\partial\mu}S_{A}^{(n)}=-\mathcal{A}\cdot\frac{c}{6}.
\end{equation}
It is quite interesting to note that this RG flow induced $n$-th R\'enyi  entropy $S_{A}^{(n)}$
is independent of $n$ and therefore simply equals the RG flow induced IR EE!     
After replacing $\mu_{\text{UV}}^{-1}/\mu_{\text{IR}}^{-1}$ by the UV/IR
cutoff $\epsilon/\xi$, the RG flow induced IR EE  is
given by
\begin{equation}
S^{\rm IR}_A= S_{A}^{(n)}=-\int_{\mu_{\text{UV}}}^{\mu_{\text{IR}}}\mathcal{A}\cdot\frac{c}{6}\frac{d\mu}{\mu}=\mathcal{A}\cdot\frac{c}{6}\log\frac{\xi}{\epsilon},
\label{eq: IREE1}
\end{equation}
which  is independent of the metric $\gamma$.

Two other field theoretic approaches to derive this IR EE (\ref{eq: IREE1}) were given in Ref. \cite{Calabrese:2004eu}. 
The first one parallels the proof of $c-$theorem. 
The second argument calculates a scalar field theory perturbed by a mass term.  
None of them requires unitarity. However, both approaches have to take the limit $n\to 1$ to get the IR EE from the 
$n$-th R\'enyi entropy, and  the number of the boundary points $\mathcal{A}$ cannot be easily derived.

The IR EE (\ref{eq: IREE1}) is universal and determined 
only by the central charge and the intrinsic correlation length of a specific CFT.
In the  $\text{dS}$/$\text{CFT}$ context, it is known that the central charge of CFT$_2$ dual to dS$_3$ is
imaginary-valued \cite{Maldacena:2002vr}, and from the Brown-Henneaux's formula \cite{Brown1986}, we have 
\begin{equation}
c=-\text{i}\,c_{\text{dS}}=-\text{i}\,\frac{3\ell_{\text{dS}}}{2G_{N}^{(3)}},
\label{eq: Brown-Henneaux}
\end{equation}
Thus, the IR EE in the $\text{dS}_{3}$/$\text{CFT}_{2}$
correspondence is
\begin{equation}
S^{\rm IR}_{A}=-\frac{\text{i}\,c_{\text{dS}}}{6}\log\frac{\xi}{\epsilon} 
= -\text{i}\,\frac{\ell_{\text{dS}}}{4G_{N}^{(3)}} \log\frac{\xi}{\epsilon},
\label{eq: IREE}
\end{equation}
where  $\mathcal{A}=1$ is assumed.

It is crucial to understand that the IR EE should not be confused with the UV EE,
\begin{equation}
S_{A}^{\text{UV}}=-\frac{\text{i}\,c_{\text{dS}}}{3}\log\frac{L}{\epsilon}+\frac{\pi c_{\text{dS}}}{6}.
\end{equation}
The differences are quite distinct from the above expressions, since the IR EE does not 
have a real part but the UV EE does\footnote{This is not the case in the AdS/CFT context, where both the IR EE
and the traditional spacelike UV EE are real. 
Perhaps this is why IR EE is frequently overlooked as an independent EE, but
mistakenly considered as only one half of the UV EE.}. 
Additionally, the IR cutoff $\xi$ used in the IR EE
has a clear different interpretation from the  entangling interval length $L=\Delta x$ used in the UV EE.
The UV EE only holds as $L << \xi$.

%

As explained in the Introduction, in the dS/CFT context, the RG flow in 
the CFT corresponds the temporal direction of dS, thus the RG flow induced IR EE
is nothing but a timelike EE in dS/CFT.

\section{Holographic timelike entanglement entropy in $\text{dS}_{3}$/$\text{CFT}_{2}$}


To exhibit the timelike feature of the IR EE  explicitly, 
it is illuminating to study its corresponding dS bulk dual. 
For any two points $(t_1, x_1, y_1)$ and $(t_2, x_2, y_2)$  
in the planar coordinate of $\text{dS}_{3}$ (\ref{eq:flat_metric}),
the  geodesic distance $L$ is
\begin{equation}
\cos\left(\frac{L}{\ell_{\text{dS}}}\right)=1+\frac{\left[\exp\left(-t_{1} /\ell_{\text{dS}}
\right)-\exp\left(-t_{2}/\ell_{\text{dS}}\right)\right]^{2}-
\frac{1}{\ell^2_{\text{dS}}}\left[\left(x_{1}-x_{2}\right)^{2}+\left(y_{1}-y_{2}\right)^{2}\right]}
{2\exp\left(-t_{1}/\ell_{\text{dS}}\right)\exp\left(-t_{2}/\ell_{\text{dS}}\right)}.
\end{equation}
The length of
a timelike geodesic between two points $\left(t_{+\infty}=\ell_{\text{dS}}\log(\ell_{\text{dS}}/\epsilon),x,y\right)$
and $\left(t_{-\infty}=\ell_{\text{dS}}\log(\ell_{\text{dS}}/\xi),x,y\right)$
is {\it exactly}
\[
L\left(t_{+\infty},t_{-\infty}\right)=\ell_{\text{dS}}\arccos\left[\frac{\xi^{2}+\epsilon^{2}}{2\xi\epsilon}\right]=-\text{i}\,\ell_{\text{dS}}\log\left(\frac{\xi}{\epsilon}\right).
\]
where the principal branch of the complex inverse cosine function is chosen. 
Assuming the Ryu-Takayanagi formula \cite{Ryu:2006bv} also holds in dS/CFT,
applying 
eqn. (\ref{eq: Brown-Henneaux}),
this timelike geodesic length gives the corresponding entropy  
\begin{equation}
S_{A} = \frac{L}{4G_N^{(3)}} =-\frac{\text{i}\,\ell_{\text{dS}}}{4G^{(3)}_{N}}\log\frac{\xi}{\epsilon}=-\frac{\text{i}\,c_{\text{dS}}}{6}\log\frac{\xi}{\epsilon}.
\end{equation}
which is {\it precisely} equal to the IR EE in  eqn. (\ref{eq: IREE}). So, we
do find perfectly matched quantities in dS$_3$ and CFT$_2$. One is 
a timelike geodesic, another is the IR EE. Intriguingly, the match is exact and there is
no need to take the UV or IR limits. 

It is particularly evident that, under the double Wick rotation (\ref{eq: Wick rotation}), 
the timelike IR EE in dS/CFT is transformed to a spacelike IR EE in AdS/CFT. 
In both dS and AdS, 
the geodesics that are dual to the IR EE extend all the way into the bulk, in sharp contrast to the UV EE 
whose endpoints are both attached to the boundary. 
As a result, the IR EE must provide indispensable information for the reconstruction of spacetime.

The various EEs in dS and AdS are connected via analytic continuations. 
In the planar or Poincare coordinates, they are transformed to each other  through  
the double Wick rotation (\ref{eq: Wick rotation}). 
However, this procedure is far from  clear
beyond the semiclassical limit and some more nontrivial calculation may be needed.
We summarize the classifications of all the EEs in 
Table (\ref{tab:1}).

\begin{table}
\begin{onehalfspace}
\begin{centering}
\begin{tabular}{|c|l|l|}
\cline{2-3} \cline{3-3} 
\multicolumn{1}{c|}{} & dS/CFT & AdS/CFT\tabularnewline
\hline 
\multirow{2}{*}{UV} & Spacelike: $S_{A}=-\frac{\text{i}\,c_{\text{dS}}}{3}\log\left(\frac{X}{\epsilon}\right)+\frac{\pi c_{\text{dS}}}{6}$ & Spacelike: $S_{A}=\frac{c_{\text{AdS}}}{3}\log\frac{X}{\epsilon}$\tabularnewline
\cline{2-3} \cline{3-3} 
 & Spacelike: $S_{A}=-\frac{\text{i}\,c_{\text{dS}}}{3}\log\left(\frac{Y}{\epsilon}\right)+\frac{\pi c_{\text{dS}}}{6}$ & $\,$Timelike: $S_{A}=\frac{c_{\text{AdS}}}{3}\log\frac{T}{\epsilon}+\frac{\text{i}\,\pi c_{\text{AdS}}}{6}$\tabularnewline
\hline 
IR & $\,$Timelike: $S_{A}=-\frac{\text{i}\,c_{\text{dS}}}{6}\mathcal{A} \log\frac{\xi}{\epsilon}$ & Spacelike: $S_{A}=\frac{c_{\text{AdS}}}{6} \mathcal{A}\log\frac{\xi}{\epsilon}$\tabularnewline
\hline 
\end{tabular}
\par\end{centering}
\end{onehalfspace}
\caption{Spacelike, timelike EEs and UV, IR EEs in  dS/CFT and AdS/CFT.  
Two elements in each row are transformed to each other via an analytic continuation.
Evidently, 
the IR EE and UV EE are completely distinct, especially from the perspective of dS/CFT.
$\mathcal{A}$ is the number of the boundary points of the subsystem $A$.}
\label{tab:1}
\end{table}


\section{Conclusions}

\label{Sec:Conc}

In this paper, we introduced a timelike EE,  in the context of
dS/CFT correspondence. Since this RG flow induced EE is expressed
by both IR and UV cutoffs, we called it as IR EE to distinguish it from the
usual UV EE, which involves UV cutoff only. 
We demonstrated that this  IR EE does perfectly match the length of a timelike
geodesic connecting two distinct spacelike surfaces in dS$_3$. 
In AdS$_3$, the counterpart of this IR EE is spacelike.
Our results reveal that there are three independent EEs in whether dS$_3$/CFT$_2$
or  AdS$_3$/CFT$_2$, which  provides just enough  
information to reconstruct the bulk geometry.

It is quite intriguing that, the match of this IR EE with the dual geodesic
length is exact, working for any cutoffs, not restricted to the UV or IR limits.

While we considered
the simplest pure dS in this paper, our findings could be extended
to generic asymptotic dS(AdS) spacetimes whose time(radial) evolution corresponds
to a nontrivial inverse RG flow. 

It is difficult to ignore the potential role of the de Sitter
cosmological event horizon as a natural entangling surface. 
It is very interesting that  there is one candidate, namely the $T\bar T$ deformation,
could realize this idea.

Refer to Figure (\ref{fig:Penrose}),  
the UV and the IR are connected at $r=\infty$, which leads us to
speculate that the timelike EE, represented by
the  green dashed lines, may serve as a holographic screen
for the inflation patch of de Sitter spacetime. This suggests that
the inflation patch is in a mixed state and is entangled with another
universe, or possibly even multiple universes.

%

\vspace{3ex} 
\noindent {\bf Acknowledgements} 
This work is supported in part by NSFC (Grant No. 12275183, 12275184,12105191 and 11875196). HW is 
partly supported by discussions arising from the DAMTP workshop "Quantum de Sitter Universe", funded by the Gravity Theory Trust and the Centre for Theoretical Cosmology. HW is also supported by the International Visiting Program for Excellent Young Scholars of Sichuan University.

\appendix

\bibliographystyle{unsrturl}
\bibliography{dS_IR_v3}

\begin{thebibliography}{10}

\bibitem{Maldacena:1997re}
Juan~Martin Maldacena.
\newblock {The Large N limit of superconformal field theories and
  supergravity}.
\newblock {\em Adv. Theor. Math. Phys.}, 2:231--252, 1998.
\newblock \href {http://arxiv.org/abs/hep-th/9711200}
  {\path{arXiv:hep-th/9711200}}, \href
  {https://doi.org/10.1023/A:1026654312961}
  {\path{doi:10.1023/A:1026654312961}}.

\bibitem{Witten:1998qj}
Edward Witten.
\newblock {Anti-de Sitter space and holography}.
\newblock {\em Adv. Theor. Math. Phys.}, 2:253--291, 1998.
\newblock \href {http://arxiv.org/abs/hep-th/9802150}
  {\path{arXiv:hep-th/9802150}}, \href
  {https://doi.org/10.4310/ATMP.1998.v2.n2.a2}
  {\path{doi:10.4310/ATMP.1998.v2.n2.a2}}.

\bibitem{Aharony:1999ti}
Ofer Aharony, Steven~S. Gubser, Juan~Martin Maldacena, Hirosi Ooguri, and Yaron
  Oz.
\newblock {Large N field theories, string theory and gravity}.
\newblock {\em Phys. Rept.}, 323:183--386, 2000.
\newblock \href {http://arxiv.org/abs/hep-th/9905111}
  {\path{arXiv:hep-th/9905111}}, \href
  {https://doi.org/10.1016/S0370-1573(99)00083-6}
  {\path{doi:10.1016/S0370-1573(99)00083-6}}.

\bibitem{Strominger:2001pn}
Andrew Strominger.
\newblock {The dS / CFT correspondence}.
\newblock {\em JHEP}, 10:034, 2001.
\newblock \href {http://arxiv.org/abs/hep-th/0106113}
  {\path{arXiv:hep-th/0106113}}, \href
  {https://doi.org/10.1088/1126-6708/2001/10/034}
  {\path{doi:10.1088/1126-6708/2001/10/034}}.

\bibitem{Bousso:2002ju}
Raphael Bousso.
\newblock {The Holographic principle}.
\newblock {\em Rev. Mod. Phys.}, 74:825--874, 2002.
\newblock \href {http://arxiv.org/abs/hep-th/0203101}
  {\path{arXiv:hep-th/0203101}}, \href
  {https://doi.org/10.1103/RevModPhys.74.825}
  {\path{doi:10.1103/RevModPhys.74.825}}.

\bibitem{Ryu:2006bv}
Shinsei Ryu and Tadashi Takayanagi.
\newblock {Holographic derivation of entanglement entropy from AdS/CFT}.
\newblock {\em Phys. Rev. Lett.}, 96:181602, 2006.
\newblock \href {http://arxiv.org/abs/hep-th/0603001}
  {\path{arXiv:hep-th/0603001}}, \href
  {https://doi.org/10.1103/PhysRevLett.96.181602}
  {\path{doi:10.1103/PhysRevLett.96.181602}}.

\bibitem{Ryu:2006ef}
Shinsei Ryu and Tadashi Takayanagi.
\newblock Aspects of holographic entanglement entropy.
\newblock {\em Journal of High Energy Physics}, 2006(08):045--045, aug 2006.
\newblock \href {https://doi.org/10.1088/1126-6708/2006/08/045}
  {\path{doi:10.1088/1126-6708/2006/08/045}}.

\bibitem{Wang:2018jva}
Peng Wang, Houwen Wu, and Haitang Yang.
\newblock {Fix the dual geometries of $T\bar{T}$ deformed CFT$_2$ and highly
  excited states of CFT$_2$}.
\newblock {\em Eur. Phys. J. C}, 80(12):1117, 2020.
\newblock \href {http://arxiv.org/abs/1811.07758} {\path{arXiv:1811.07758}},
  \href {https://doi.org/10.1140/epjc/s10052-020-08680-7}
  {\path{doi:10.1140/epjc/s10052-020-08680-7}}.

\bibitem{Doi:2022iyj}
Kazuki Doi, Jonathan Harper, Ali Mollabashi, Tadashi Takayanagi, and Yusuke
  Taki.
\newblock {Pseudoentropy in dS/CFT and Timelike Entanglement Entropy}.
\newblock {\em Phys. Rev. Lett.}, 130(3):031601, 2023.
\newblock \href {http://arxiv.org/abs/2210.09457} {\path{arXiv:2210.09457}},
  \href {https://doi.org/10.1103/PhysRevLett.130.031601}
  {\path{doi:10.1103/PhysRevLett.130.031601}}.

\bibitem{Narayan:2022afv}
K.~Narayan.
\newblock {de Sitter space, extremal surfaces and ''time-entanglement''}.
\newblock 10 2022.
\newblock \href {http://arxiv.org/abs/2210.12963} {\path{arXiv:2210.12963}}.

\bibitem{Li:2022tsv}
Ze~Li, Zi-Qing Xiao, and Run-Qiu Yang.
\newblock {On holographic time-like entanglement entropy}.
\newblock 11 2022.
\newblock \href {http://arxiv.org/abs/2211.14883} {\path{arXiv:2211.14883}}.

\bibitem{Doi:2023zaf}
Kazuki Doi, Jonathan Harper, Ali Mollabashi, Tadashi Takayanagi, and Yusuke
  Taki.
\newblock {Timelike entanglement entropy}.
\newblock 2 2023.
\newblock \href {http://arxiv.org/abs/2302.11695} {\path{arXiv:2302.11695}}.

\bibitem{Jiang:2023ffu}
Xin Jiang, Peng Wang, Houwen Wu, and Haitang Yang.
\newblock {Timelike entanglement entropy and $T\bar{T}$ deformation}.
\newblock 2 2023.
\newblock \href {http://arxiv.org/abs/2302.13872} {\path{arXiv:2302.13872}}.

\bibitem{Narayan:2023ebn}
K.~Narayan and Hitesh~K. Saini.
\newblock {Notes on time entanglement and pseudo-entropy}.
\newblock 3 2023.
\newblock \href {http://arxiv.org/abs/2303.01307} {\path{arXiv:2303.01307}}.

\bibitem{Maldacena:2002vr}
Juan~Martin Maldacena.
\newblock {Non-Gaussian features of primordial fluctuations in single field
  inflationary models}.
\newblock {\em JHEP}, 05:013, 2003.
\newblock \href {http://arxiv.org/abs/astro-ph/0210603}
  {\path{arXiv:astro-ph/0210603}}, \href
  {https://doi.org/10.1088/1126-6708/2003/05/013}
  {\path{doi:10.1088/1126-6708/2003/05/013}}.

\bibitem{Hikida:2021ese}
Yasuaki Hikida, Tatsuma Nishioka, Tadashi Takayanagi, and Yusuke Taki.
\newblock {Holography in de Sitter Space via Chern-Simons Gauge Theory}.
\newblock {\em Phys. Rev. Lett.}, 129(4):041601, 2022.
\newblock \href {http://arxiv.org/abs/2110.03197} {\path{arXiv:2110.03197}},
  \href {https://doi.org/10.1103/PhysRevLett.129.041601}
  {\path{doi:10.1103/PhysRevLett.129.041601}}.

\bibitem{Hikida:2022ltr}
Yasuaki Hikida, Tatsuma Nishioka, Tadashi Takayanagi, and Yusuke Taki.
\newblock {CFT duals of three-dimensional de Sitter gravity}.
\newblock {\em JHEP}, 05:129, 2022.
\newblock \href {http://arxiv.org/abs/2203.02852} {\path{arXiv:2203.02852}},
  \href {https://doi.org/10.1007/JHEP05(2022)129}
  {\path{doi:10.1007/JHEP05(2022)129}}.

\bibitem{Strominger:2001gp}
Andrew Strominger.
\newblock {Inflation and the dS / CFT correspondence}.
\newblock {\em JHEP}, 11:049, 2001.
\newblock \href {http://arxiv.org/abs/hep-th/0110087}
  {\path{arXiv:hep-th/0110087}}, \href
  {https://doi.org/10.1088/1126-6708/2001/11/049}
  {\path{doi:10.1088/1126-6708/2001/11/049}}.

\bibitem{Wang:2017bym}
Peng Wang, Houwen Wu, and Haitang Yang.
\newblock {AdS$_3$ metric from UV/IR entanglement entropies of CFT$_2$}.
\newblock 10 2017.
\newblock \href {http://arxiv.org/abs/1710.08448} {\path{arXiv:1710.08448}}.

\bibitem{Wang:2018vbw}
Peng Wang, Houwen Wu, and Haitang Yang.
\newblock {Derive three dimensional geometries from entanglement entropies of
  CFT$_2$}.
\newblock 9 2018.
\newblock \href {http://arxiv.org/abs/1809.01355} {\path{arXiv:1809.01355}}.

\bibitem{Anninos:2011ui}
Dionysios Anninos, Thomas Hartman, and Andrew Strominger.
\newblock {Higher Spin Realization of the dS/CFT Correspondence}.
\newblock {\em Class. Quant. Grav.}, 34(1):015009, 2017.
\newblock \href {http://arxiv.org/abs/1108.5735} {\path{arXiv:1108.5735}},
  \href {https://doi.org/10.1088/1361-6382/34/1/015009}
  {\path{doi:10.1088/1361-6382/34/1/015009}}.

\bibitem{Narayan:2015vda}
K.~Narayan.
\newblock {Extremal surfaces in de Sitter spacetime}.
\newblock {\em Phys. Rev. D}, 91(12):126011, 2015.
\newblock \href {http://arxiv.org/abs/1501.03019} {\path{arXiv:1501.03019}},
  \href {https://doi.org/10.1103/PhysRevD.91.126011}
  {\path{doi:10.1103/PhysRevD.91.126011}}.

\bibitem{Sato2015}
Yoshiki Sato.
\newblock Comments on entanglement entropy in the ds/cft correspondence.
\newblock {\em Phys. Rev. D}, 91(8):086009, 2015.
\newblock \href {http://arxiv.org/abs/1501.04903} {\path{arXiv:1501.04903}},
  \href {https://doi.org/10.1103/PhysRevD.91.086009}
  {\path{doi:10.1103/PhysRevD.91.086009}}.

\bibitem{Narayan:2015oka}
K.~Narayan.
\newblock {de Sitter space and extremal surfaces for spheres}.
\newblock {\em Phys. Lett. B}, 753:308--314, 2016.
\newblock \href {http://arxiv.org/abs/1504.07430} {\path{arXiv:1504.07430}},
  \href {https://doi.org/10.1016/j.physletb.2015.12.019}
  {\path{doi:10.1016/j.physletb.2015.12.019}}.

\bibitem{Narayan:2016xwq}
K.~Narayan.
\newblock {On $dS_4$ extremal surfaces and entanglement entropy in some ghost
  CFTs}.
\newblock {\em Phys. Rev. D}, 94(4):046001, 2016.
\newblock \href {http://arxiv.org/abs/1602.06505} {\path{arXiv:1602.06505}},
  \href {https://doi.org/10.1103/PhysRevD.94.046001}
  {\path{doi:10.1103/PhysRevD.94.046001}}.

\bibitem{Nakata:2020luh}
Yoshifumi Nakata, Tadashi Takayanagi, Yusuke Taki, Kotaro Tamaoka, and Zixia
  Wei.
\newblock {New holographic generalization of entanglement entropy}.
\newblock {\em Phys. Rev. D}, 103(2):026005, 2021.
\newblock \href {http://arxiv.org/abs/2005.13801} {\path{arXiv:2005.13801}},
  \href {https://doi.org/10.1103/PhysRevD.103.026005}
  {\path{doi:10.1103/PhysRevD.103.026005}}.

\bibitem{Mollabashi:2020yie}
Ali Mollabashi, Noburo Shiba, Tadashi Takayanagi, Kotaro Tamaoka, and Zixia
  Wei.
\newblock {Pseudo Entropy in Free Quantum Field Theories}.
\newblock {\em Phys. Rev. Lett.}, 126(8):081601, 2021.
\newblock \href {http://arxiv.org/abs/2011.09648} {\path{arXiv:2011.09648}},
  \href {https://doi.org/10.1103/PhysRevLett.126.081601}
  {\path{doi:10.1103/PhysRevLett.126.081601}}.

\bibitem{Nishioka:2021cxe}
Tatsuma Nishioka, Tadashi Takayanagi, and Yusuke Taki.
\newblock {Topological pseudo entropy}.
\newblock {\em JHEP}, 09:015, 2021.
\newblock \href {http://arxiv.org/abs/2107.01797} {\path{arXiv:2107.01797}},
  \href {https://doi.org/10.1007/JHEP09(2021)015}
  {\path{doi:10.1007/JHEP09(2021)015}}.

\bibitem{Mollabashi:2021xsd}
Ali Mollabashi, Noburo Shiba, Tadashi Takayanagi, Kotaro Tamaoka, and Zixia
  Wei.
\newblock {Aspects of pseudoentropy in field theories}.
\newblock {\em Phys. Rev. Res.}, 3(3):033254, 2021.
\newblock \href {http://arxiv.org/abs/2106.03118} {\path{arXiv:2106.03118}},
  \href {https://doi.org/10.1103/PhysRevResearch.3.033254}
  {\path{doi:10.1103/PhysRevResearch.3.033254}}.

\bibitem{Calabrese:2004eu}
Pasquale Calabrese and John~L. Cardy.
\newblock {Entanglement entropy and quantum field theory}.
\newblock {\em J. Stat. Mech.}, 0406:P06002, 2004.
\newblock \href {http://arxiv.org/abs/hep-th/0405152}
  {\path{arXiv:hep-th/0405152}}, \href
  {https://doi.org/10.1088/1742-5468/2004/06/P06002}
  {\path{doi:10.1088/1742-5468/2004/06/P06002}}.

\bibitem{Calabrese:2009qy}
Pasquale Calabrese and John Cardy.
\newblock {Entanglement entropy and conformal field theory}.
\newblock {\em J. Phys. A}, 42:504005, 2009.
\newblock \href {http://arxiv.org/abs/0905.4013} {\path{arXiv:0905.4013}},
  \href {https://doi.org/10.1088/1751-8113/42/50/504005}
  {\path{doi:10.1088/1751-8113/42/50/504005}}.

\bibitem{Osborn:1991gm}
H.~Osborn.
\newblock {Weyl consistency conditions and a local renormalization group
  equation for general renormalizable field theories}.
\newblock {\em Nucl. Phys. B}, 363:486--526, 1991.
\newblock \href {https://doi.org/10.1016/0550-3213(91)80030-P}
  {\path{doi:10.1016/0550-3213(91)80030-P}}.

\bibitem{Fursaev:1995ef}
Dmitri~V. Fursaev and Sergey~N. Solodukhin.
\newblock {On the description of the Riemannian geometry in the presence of
  conical defects}.
\newblock {\em Phys. Rev. D}, 52:2133--2143, 1995.
\newblock \href {http://arxiv.org/abs/hep-th/9501127}
  {\path{arXiv:hep-th/9501127}}, \href
  {https://doi.org/10.1103/PhysRevD.52.2133}
  {\path{doi:10.1103/PhysRevD.52.2133}}.

\bibitem{Brown1986}
J.~David Brown and M.~Henneaux.
\newblock Central charges in the canonical realization of asymptotic
  symmetries: An example from three-dimensional gravity.
\newblock {\em Commun. Math. Phys.}, 104:207--226, 1986.
\newblock \href {https://doi.org/10.1007/BF01211590}
  {\path{doi:10.1007/BF01211590}}.

\end{thebibliography}

\end{document}